\documentclass{sig-alternate-per}

\usepackage{booktabs}
\usepackage{marvosym}

\begin{document}
%
\conferenceinfo{W-PIN 2012}{London, UK}

\title{Pricing of insurance policies against cloud storage price rises}
\subtitle{[Extended Abstract]}
%
%
%
%
%

\numberofauthors{2} 
%
\author{
%
%
\alignauthor
Loretta Mastroeni\\
       \affaddr{Universit\`{a} di Roma Tre}\\
       \affaddr{Dip. di Economia}\\
       \affaddr{Rome, Italy}\\
       \email{mastroen@uniroma3.it}
\alignauthor
Maurizio Naldi\\
       \affaddr{Universit\`{a} di Roma Tor Vergata}\\
       \affaddr{Dip. di Informatica Sistemi Produzione}\\
       \affaddr{Rome, Italy}\\
       \email{naldi@disp.uniroma2.it}
}
\date{30 July 1999}

\maketitle
\begin{abstract}
When a company migrates to cloud storage, the way back is neither easy nor cheap. The company is then locked up in the storage contract and exposed to upward market prices, which reduce the company's profit and may even bring it below zero. We propose a protection means based on an insurance contract, by which the cloud purchaser is indemnified when the current storage price exceeds a pre-defined threshold. By applying the financial options theory, we provide a formula for the insurance price (the premium). By using historical data on market prices for disks, we apply the formula in realistic scenarios. We show that the premium grows nearly quadratically with the length of the coverage period as long as this is below one year, but grows more slowly, though faster than linearly, over longer coverage periods.
\end{abstract}



\section{Introduction}
Cloud storage is a major component of the virtualization process. In cloud storage, a customer stores its data on the  facilities of a cloud provider and can then access them through the Internet.  The cloud replaces the customer's data infrastructure and allows the cloud purchaser to pursue an outsourcing strategy: it is a major example of the IaaS (Infrastructure as a Service) paradigm. 

Cloud storage service is paid for though a periodic fee related to the amount of data stored. Several papers have been devoted to the economical analysis of the migration from an owned infrastructure to the cloud \cite{Greenwood, Walker10, NaldiCEC2011}. The profitability of migration relies heavily on the price required by the cloud provider: too high a price erodes the profit margins and may make infrastructure ownership the best business proposition. Since the migration decision must be evaluated in the long run \cite{NaldiCEC2011}, any decision has to be based on  forecasts for future prices. Given the long evaluation window, there is a significant probability of taking the wrong decision. In \cite{NaldiCEC2011}, the Value-at-Risk metric was employed to analyse the extent of such risk. The cloud purchaser has to deal with such a risk and look for mitigating strategies.

In \cite{Princehouse10} a technology-based strategy has been proposed to reduce the problems due to price hikes, through the application of RAID-like techniques (\textit{Redundant Array of Independent Disks}). That strategy is not free, since it involves managing multiple storage contracts and the extra storage costs to achieve redundancy.

Here we propose an insurance-based strategy, which may be employed as either an alternative or an accompanying strategy for any technology-based solution. It calls for the cloud purchaser underwriting an insurance policy to hedge against price increases. Actually, since a downward trend is expected for cloud storage prices, our insurance approach achieves protection against any upward deviation from the downward trend. We envisage an insurance contract covering a long-term migration strategy. Since a major issue is the cost associated to the strategy, we  derive a pricing formula for the insurance contract and evaluate the resulting price for several realistic scenarios. We show that the insurance price (the \textit{premium}) grows nearly quadratically with the length of the coverage period as long as this is below one year, but grows more slowly, though faster than linearly, over longer coverage periods.

The paper is organized as follows. In Section \ref{fluct}, we review the current cloud storage market and describe a model for future price fluctuations.  The insurance policy and the pricing formula are described in Section \ref{Insur}. In Section \ref{sample}, we set some sample scenarios and evaluate the resulting premium.

\section{Price fluctuations}
\label{fluct}
In order to get established as an operational paradigm, cloud storage must be economically profitable both for the cloud provider and for the cloud purchaser. Costs for the cloud purchaser are determined by the prices offered by the cloud provider, which in turn are lower bounded by its costs. Such costs pose different constraints on the two parties. Cloud providers must set prices at least so high as to recover their costs: costs represent a lower bound for prices. Instead, cloud purchasers seek prices lower than their costs (otherwise they would stay with their data center): costs represent an upper bound for prices. Whatever the role of the party, we expect costs to vary over time, under the influence of a number of factors (e.g., cost of disks, operational expenses, level of competition). Likewise, we expect prices to reflect those changes and vary as well. We can model the fluctuations of price through a stochastic model. In this section, starting from data derived from the market of disks and storage facilities, we describe the stochastic model we employ for future cloud storage prices. 

All the most important providers charge a fixed price for each time period (either a month or a year), with a maximum amount of storage space. (increased in brackets). We have examined the price plans offered by major cloud providers, which offer separate plans for business customers and consumers, one of the main differences being that consumers can access their data from just one computer. Since the unit price decreases as the amount of stored data increases, we have taken the least cost for each provider, i.e., that corresponding to the maximum  amount of data. The results of this analysis are summarized in Table \ref{table:avprice}, the prices being expressed in euros per GB and per month (prices in US dollars have been converted into euros through a conversion rate of 1.3). Prices show a significant dispersion around their average value, though we must recognize that the price itself is just one of the characteristics of the service bundle: different prices may reflect quite different service features. \begin{table}
\begin{center}
\caption{Unit prices of cloud providers }
\label{table:avprice}
\begin{tabular}{lcc}
\toprule
Provider & \multicolumn{2}{c}{Unit price [\EUR]}\\
\cmidrule{2-3}
 & Consumers & Business\\
\midrule
Average & 0.0955 & 0.185\\
Standard deviation & 0.0663 & 0.145\\
\bottomrule
\end{tabular}
\end{center}
\end{table}

The significant dispersion may be due to the infancy of the storage service, where cloud providers have quite different backgrounds and operational strategies. However, they are just present prices and tell us very little about the future trends of prices. Hence, they do not provide enough information to evaluate the profitability of the cloud approach, especially for the typically long range adopted to evaluate the migration to cloud storage as in \cite{NaldiCEC2011}. 

For that purpose, we need a forecast of future storage prices. Unfortunately, the time series of cloud storage prices are not long and rich enough to draw conclusions. More data are instead available on market prices for disks. In fact, we can use the data coming from an extensive survey conducted on SATA (\textit{Serial Advanced Technology Attachment}) disk prices every week over more than 5 years. That survey has shown that the price $P(t)$ of disks follows a decaying trend, which is approximately exponential \cite{Walker10}:
\begin{equation}
\label{pricereg}
P(t)\simeq P_{0}e^{-\beta t},
\end{equation} 
where $P_{0}$ is the price at the beginning of the observation period, $t$ is the time expressed in years, and $\beta =0.438$.

In the survey reported in \cite{Walker10}, Equation (\ref{pricereg}) was taken as a deterministic one. Actually, that equation was obtained by regression, since real prices fluctuate around the values predicted by that exponential curve. In \cite{NaldiNGI11} we have abandoned the deterministic model, assuming instead that the future price is described by a Geometric Brownian Motion (GBM) random process \cite{Chan}, where the price at time $t$ is
\begin{equation}
\label{gbm}
P(t)=P_{0}e^{(-\beta - \sigma^{2}/2)t + \sigma W(t)},
\end{equation}
with $W(t)$ being a Wiener process, and $\sigma^{2}$ the variance of the process. Since cloud providers have to purchase storage facilities and operate them, we expect their costs (and likewise their prices) to follow a process similar to that described by Equation (\ref{gbm}). We adopt therefore the Geometric Brownian Motion as a suitable model for future cloud storage prices. 

\section{Insurance coverage}
\label{Insur}
In Section \ref{fluct}, we have set a model to describe price fluctuations. Though we expect prices to go downward, they may be higher than expected. A purchaser of storage space has to protect itself against upward fluctuations. It can do so by buying an insurance policy. In this section, we formulate the insurance policy as a multiperiod call option, and derive the fair premium for such insurance policy.

As shown in Section \ref{fluct}, prices of cloud storage can go above and below their expected (negative exponential) trend. That trend is what the cloud storage purchaser expects to pay in the future. If the actual price is lower than what is predicted by the exponential curve (\ref{pricereg}), the convenience margins are higher than expected. Instead, in the case of upward fluctuations of prices, the profit margins shrink and may get negative: upward fluctuations represent a risk for the cloud purchaser. We can define the risk $R(t)$ incurred at time $t$ for each unit of storage (e.g., a GB) as the difference between the actual unit price $S(t)$ and that predicted by Equation (\ref{pricereg})
\begin{equation}
\label{risk}
R(t)=\left(S(t)-\mathbb{E}[P(t)]\right)\vert_{+}.
\end{equation}  

The cloud purchaser seeks protection against that risk. A suitable insurance policy may consist in the payment to the insurer of a sum equal to $R(t)$ whenever the actual price exceeds the expected one, against a payment by the insured (the cloud purchaser) of an initial premium (the price of the insurance contract) when the insurance contract is underwritten. By means of this policy, the cloud purchaser is completely indemnified against any price increase above the expected value $\mathbb{E}[P(t)]$. The main problem is now the computation of the fair value $\mathcal{V}$ of the premium that the cloud purchaser has to pay to buy the insurance policy (at the time conventionally set equal to zero). We have also to define when the comparison represented by Equation (\ref{risk}) is performed to assess if the insured has the right to receive the compensation. A natural solution is to use the same time period employed in pricing plans (e.g., every month). If we consider the timing at which the purchase contract is renewed (e.g., each month), we have a sequence $\{t_{1},t_{2}, \ldots , t_{n}\}$ of times when the cloud purchaser renews its contract with the cloud provider and compares the actual price with the expected one. If the actual price is larger than the expected one, the insured (the cloud purchaser) obtains the difference from the insurer. The length $n$ of the sequence is determined by the insurance contract. The sequence of cash flows for the cloud purchaser is
\begin{equation}
\label{cashflow}
C_{i} = \left\{ \begin{array}{ll}
-\mathcal{V} & \textrm{if $i=0$}\\ R(t_{i}) & \textrm{if $i=1, 2, \ldots , n$}
\end{array} \right. 
\end{equation}

Each individual cash flow falling at time $t_{i}$ is exactly what we would obtain with a call option. Call options are financial contracts whose reward for their holder is determined by the difference between the actual price of an underlying asset and a predefined exercise price \cite{Hull}. In practice, if the actual price is larger than the exercise price, the option holder gets the difference. In our case, the underlying asset is the cloud storage service, the exercise price is the expected price $\mathbb{E}[P(t)]=P_{0}e^{-\beta t}$, and the time at which the option can be exercised is each $t_{i}$. Since the cloud service contract renewal is done at a number of times $t_{i}$, the insurance contract is equivalent to a set of call options, one for each time $t_{i}$ (a multiperiod call option). As in the case of insurance contracts, a call option guarantees a right for its holder, and is therefore to be paid for. 

For a single call option, when the underlying asset price follows a GBM process, the fair price is given by the well known Black-Scholes formula. For the coverage against a price increase at the time $t_{i}$, the general expression of the fair price $C(t_{i})$ of the option (the insurance contract) is
\begin{equation}
\label{BS}
V(t_{i})=S(0)\mathbb{G}(d_{1})-K(t_{i})e^{-rt_{i}}\mathbb{G}(d_{2}),
\end{equation}
with
\begin{equation}
d_{1}=\frac{\ln \frac{S(0)}{K(t_{i})}+\left( r+\frac{\sigma^{2}}{2}\right)t_{i}}{\sigma\sqrt{t_{i}}} \quad d_{2}=\frac{\ln \frac{S(0)}{K(t_{i})}+\left( r-\frac{\sigma^{2}}{2}\right)t_{i}}{\sigma\sqrt{t_{i}}}, 
\end{equation}
where $\mathbb{G}(\cdot)$ is the standard normal distribution function, $K(\cdot)$ is the exercise price, $r$ is the risk-free interest rate, and $\sigma$ is the standard deviation of the value of the underlying asset (i.e., the cloud storage unit price).

In our case, the general expression becomes
\begin{equation}
V(t_{i})=S(0)\left[\mathbb{G}(d_{1})-e^{-(\beta +r) t_{i}}\mathbb{G}(d_{2}) \right]
\end{equation}
with
\begin{equation}
\label{constd}
d_{1}=\frac{r+\beta+\sigma^{2}/2}{\sigma}\sqrt{t_{i}} \quad
d_{2}=\frac{r+\beta-\sigma^{2}/2}{\sigma}\sqrt{t_{i}}.
\end{equation}

Finally, the overall price for the insurance contract is simply the sum of the prices pertaining to the single call options
\begin{equation}
\label{premprice}
\mathcal{V}=\sum_{i=1}^{n}V(t_{i})=S(0)\sum_{i=1}^{n}\left[\mathbb{G}(d_{1})-e^{-(\beta +r) t_{i}}\mathbb{G}(d_{2}) \right]
\end{equation}

\section{Sample insurance pricing}
\label{sample}
In Section \ref{Insur}, we have derived the pricing model for our insurance policy. The resulting price is a function of the current unit price, the risk-free interest rate, the rate of storage price decrease, and the duration of the insurance scheme. In this section, we set suitable values for these quantities and compute the insurance price for several realistic scenarios. 

We start by dealing with the standard deviation of cloud storage prices. In Section \ref{fluct}, we have analysed the current market prices for cloud storage for consumers and business customers. Here we employ the standard deviation of prices, as reported in Table \ref{table:avprice}, for $\sigma$ in Equations (\ref{constd}).

As to the risk-free interest rate, we should consider a government bond of the highest credit merit. For that purpose, we consider US Treasury bills and bonds, with a duration equal to the duration of the insurance policy. In the following, we consider two durations: 1 year, for which the interest rate is currently 0.2\%, and 5 years, for which the interest rate is currently 0.99\%.

Since the granularity that cloud providers adopt for their pricing plans is monthly (see Section \ref{fluct}), we assume that the time step in Equation (\ref{premprice}) is likewise monthly. Our insurance policy includes therefore a single payment (the premium) at time 0 and a sequence of monthly claims. 

We also note that the premium in Equation (\ref{premprice}) is proportional to the current price of cloud storage. In the following, we refer to the normalized premium $\mathcal{V}/S(0)$. It is however to be noted that the normalizing quantity $S(0)$ is the monthly price, while the premium refers to the overall duration of the insurance policy (1 or 5 years in our case). Both quantities apply, however, to the same unit of storage space. 

\begin{figure}[htbp]
\begin{center}
  \includegraphics[width=.95\columnwidth]{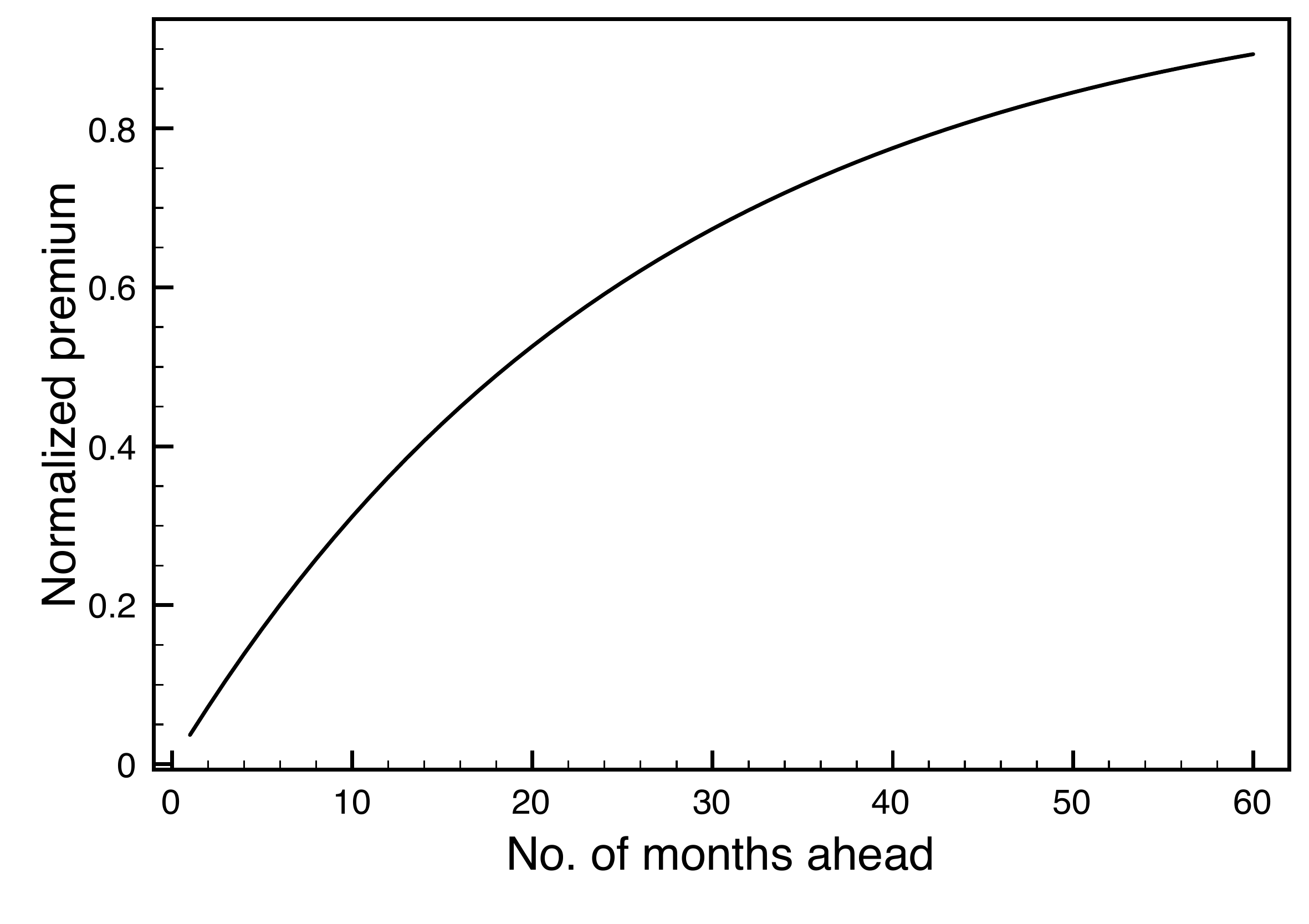}
    \caption{Normalized monthly premium over five years}
    \label{fig:p2}
\end{center}
\end{figure}

We can now compute the premium for four different cases, obtained by combining the two customer categories and two contract durations. In Table \ref{table:insprice}, we report the resulting insurance premium. We can examine the impact of several quantities intervening in the price formula (\ref{premprice}). Though the two categories of customers differ for the standard deviation of prices (with the standard deviation of prices for business customers being more than twice that for consumers), the differences in the premium to be paid look negligible, quite less than 1\%. The duration (which also brings along a variation in the risk-free interest rate) has instead a relevant role: the premium for an insurance policy covering 5 years is nearly 15 times as large as that for a single year.
\begin{table}
\begin{center}
\caption{Insurance prices}
\label{table:insprice}
\begin{tabular}{lcc}
\toprule
Category & Period [years] & $\frac{\textrm{Insurance price}}{\textrm{Current monthly price}}$ \\
\midrule
Consumer  & 1 	& 2.469 \\	
Consumer  & 5 & 36.504 \\	
Business  & 1 	& 2.481 \\	
Business & 5 & 36.516 \\	
\bottomrule
\end{tabular}
\end{center}
\end{table}
We can further analyse that issue by observing the normalized individual monthly premium $V(t_{i})$ to be paid at time 0 to cover a single monthly period $i$ months ahead. In Figure~\ref{fig:p2}, we report the resulting curve for a contract covering five years. Over the relatively short period of 1 year, the monthly premium grows roughly linearly. The premium for the first month is just 3.6\% of the current price, but grows up to 35.6\% for the premium to be paid for the twelfth month. Because of the quasi-linear growth of the monthly premium, we expect the cumulative premium $\mathcal{V}$ to grow as the sum of an arithmetic series, i.e., roughly as the square of the number of months. Instead, when we look at the trend over a longer policy duration, the growth of the monthly premium becomes quite less than linear (and correspondingly the growth of the cumulative premium will be quite less than quadratic). For the farthest month, the monthly premium reaches nearly 90\% of the current storage price.

Finally, we consider the effect of the volatility $\sigma$ of cloud storage prices and the duration of the insurance contract.

\begin{figure}[htbp]
\begin{center}
  \includegraphics[width=.95\columnwidth]{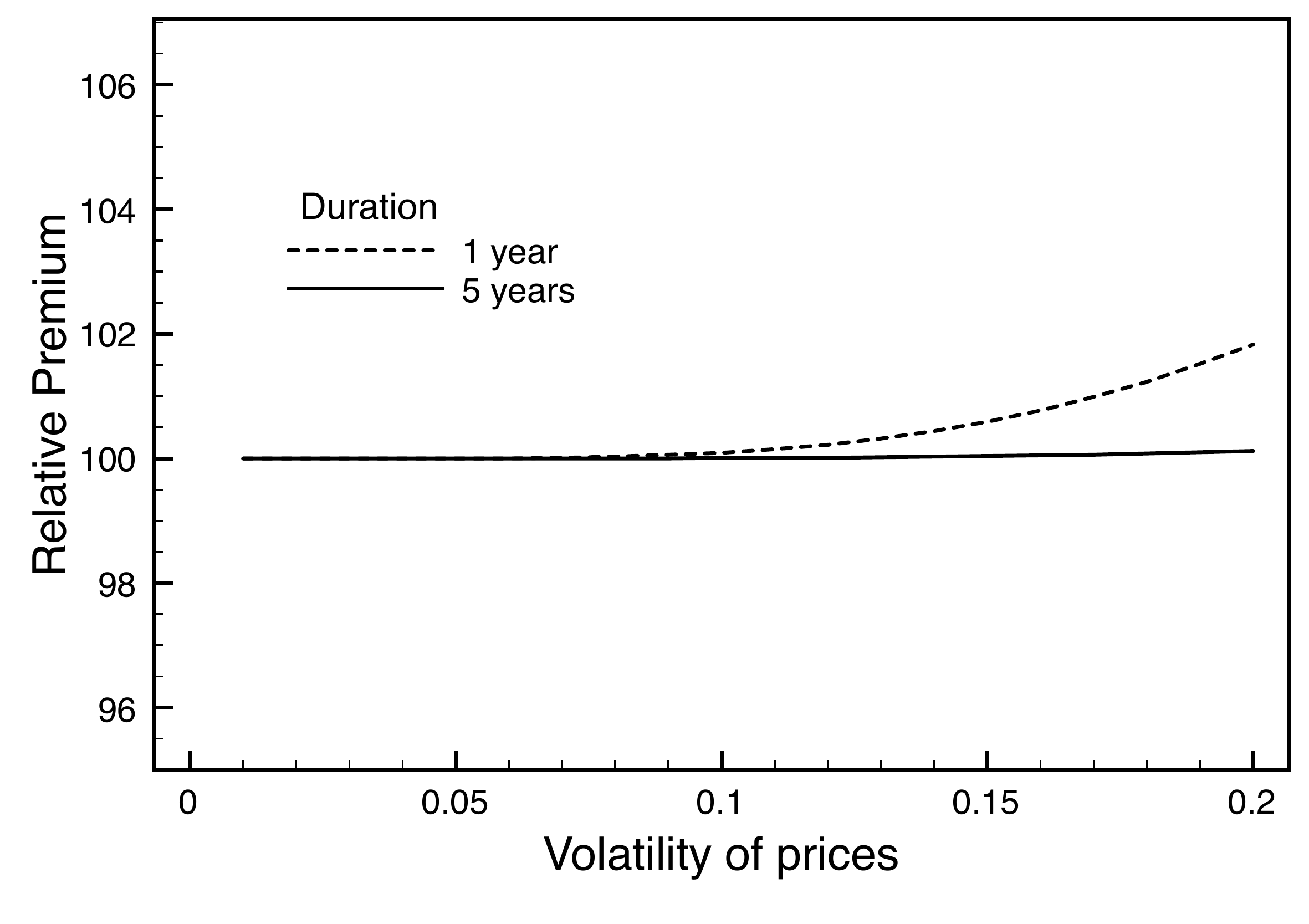}
    \caption{Impact of the volatility of prices on the premium}
    \label{fig:vareff}
\end{center}
\end{figure}

As to the former quantity, in Section \ref{fluct} we obtained an estimate for the dispersion of prices, which impacts on the premium: an incorrect estimate of the volatility would distort the insurance price. In Figure~\ref{fig:vareff}, we draw the relation between the premium and the volatility of cloud storage prices for the two contract durations considered so far, using the premium for the minimum value of the variance considered (0.01) as the baseline value. For both durations, the larger the uncertainty on the storage price (larger variance), the larger the premium. But the effect is negligible for the longer duration: over the whole range considered for the standard deviation of prices, the premium increases by 1.83\% if the duration is 1 year, but just by 0.12\% if the duration is 5 years.

\begin{figure}[htbp]
\begin{center}
  \includegraphics[width=.95\columnwidth]{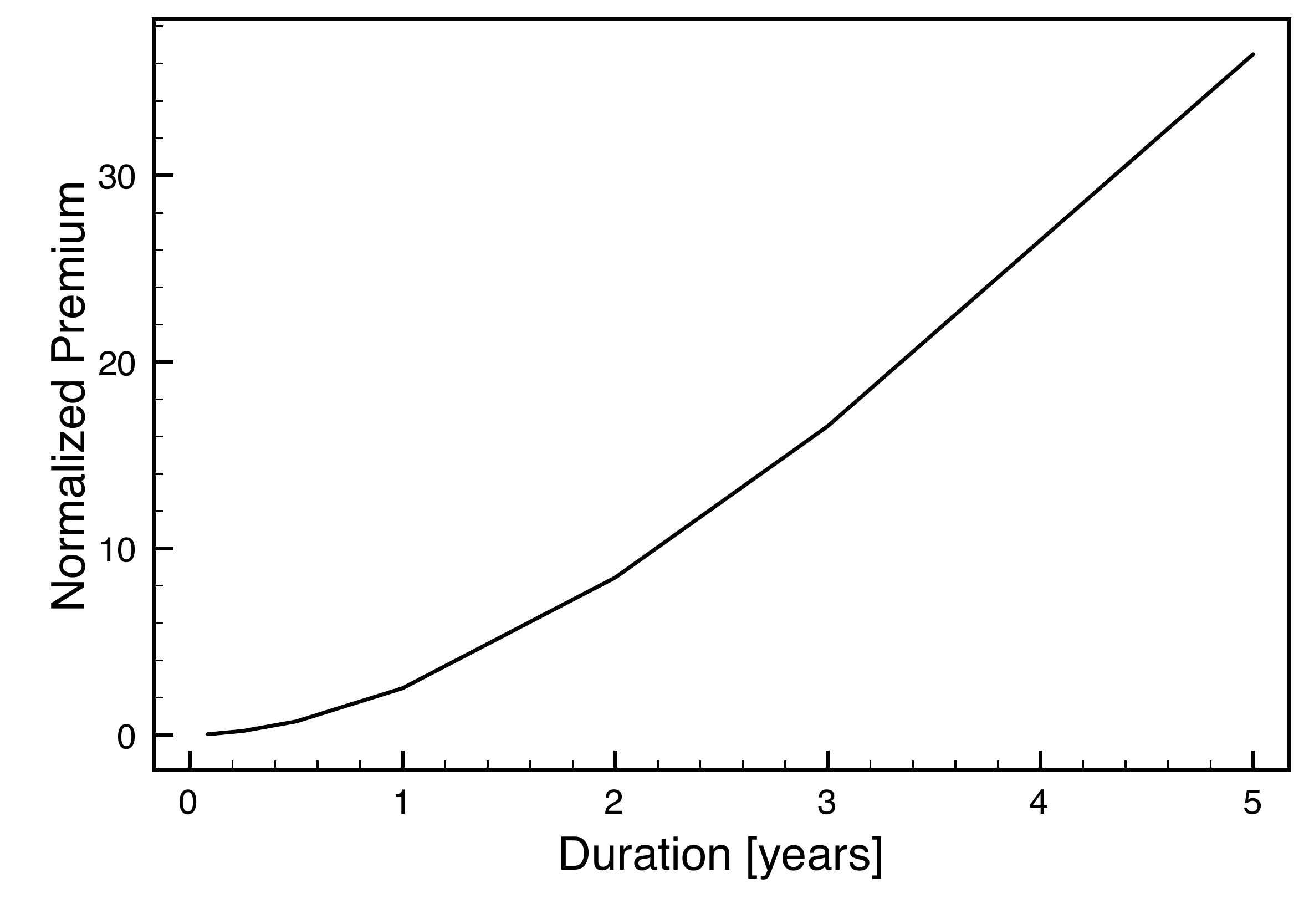}
    \caption{Impact of contract duration on the premium}
    \label{fig:dureff}
\end{center}
\end{figure}

Another relevant issue is the impact of the contract duration on the premium. In the four typical cases reported in Table \ref{table:insprice}, we observe a very large difference between the premium for 1 year and 5 years. This is due to the cumulative nature of the pricing formula (\ref{premprice}): the overall contract is actually the aggregation of a number of monthly contracts, a number that grows with the contract duration. We expect a roughly quadratic price growth for the shortest durations, and between linear and quadratic for the longer ones. In Figure~\ref{fig:dureff}, we show the actual relation between premium and duration for a typical case, where we have applied the risk-free rates pertaining to the duration. The curve confirms the  trend expected according to our above considerations.

\section{Conclusions}
Cloud storage prices are subject to changes, exposing the cloud purchaser to upward price fluctuations. But switching back from the cloud to in-house infrastructure is neither fast nor cheap, and neither is switching to another cloud provider. We have introduced an insurance scheme that protects the cloud purchaser against cloud prices exceeding the expected ones. We provide an insurance pricing formula, which delivers the premium for an insurance policy covering a sequence of monthly claims. The resulting premium has been evaluated for some realistic scenarios and grows first quadratically with the number of months and then more slowly. The premium for a single month is initially a very small fraction of the current monthly cloud storage price, but can become even 90\% of that when the covered month is 5 years from now. The impact of the dispersion of storage prices on the premium is quite small for short durations and practically negligible for longer ones.


%
%
%

\balancecolumns

\end{document}